\documentclass[fleqn]{2022AMSE_arxiv}
\usepackage{bm}

\usepackage{tabularx}
\usepackage{multirow}
\usepackage{array}
\newcolumntype{Y}{>{\raggedright\arraybackslash}X} 

\usepackage{tikz}
\usetikzlibrary{shapes.geometric}

\newcommand{\ostarfive}{%
\tikz[baseline=-0.6ex]\node[draw, star, star points=5, star point ratio=2.25,
inner sep=0pt, minimum size=1.6ex, line width=0.4pt]{};%
}

\newcommand{\ostarsix}{%
\tikz[baseline=-0.6ex]\node[draw, star, star points=6, star point ratio=2.0,
inner sep=0pt, minimum size=1.6ex, line width=0.4pt]{};%
}

\begin{document}

\ensubject{Fluid Dynamics}

\ArticleType{RESEARCH PAPER}
\Year{2027}
\Vol{43}
\DOI{10.1007/s10409-026-60186-x}
\ArtNo{360186}
\ReceiveDate{April 10, 2026}
\AcceptDate{May 6, 2026}
\OnlineDate{}

\title{Revisit eddy viscosity in pressure-driven wall turbulence at high Reynolds number}{Revisit eddy viscosity in pressure-driven wall turbulence at high Reynolds number}

\author[1]{Ben-Rui Xu}{}%
\author[1,2]{Ao Xu}{axu@nwpu.edu.cn}

\AuthorMark{B.-R. Xu}

\AuthorCitation{B.-R. Xu, and A. Xu}

\address[1]{Institute of Extreme Mechanics, School of Aeronautics, Northwestern Polytechnical University, Xi'an 710072, China}
\address[2]{National Key Laboratory of Aircraft Configuration Design, \\
Key Laboratory for Extreme Mechanics of Aircraft of Ministry of Industry and Information Technology, Xi'an 710072, China}


\abstract{\textbf{Abstract:} We investigate eddy-viscosity distributions in pressure-driven wall turbulence for three canonical configurations: plane closed-channel flow, open-channel flow with a free-slip surface, and pipe flow.
Using direct numerical simulation (DNS) databases spanning friction Reynolds numbers $Re_{\tau}=$ 2000--12000, we infer the eddy viscosity from one-point statistics through the Boussinesq relation.
The DNS-inferred eddy viscosity displays configuration-dependent behavior in the outer region, indicating that a single full-depth expression is not uniformly accurate for all three configurations.
Building on the interpretation of eddy viscosity as the product of a velocity scale and a length scale, we extend the log-law scaling into the outer region.
Specifically, we adopt a stress-based velocity scale and introduce an outer correction function to capture the remaining dependence on the outer coordinate.
We then embed a compact parametric form of this correction into a Cess-type framework with van Driest near-wall damping, yielding a full-depth eddy-viscosity model.
We assess the model using eddy-viscosity profiles, the log-law indicator function, and skin friction.
The results show that the proposed model yields noticeable improvement for open-channel flow while remaining comparable to the classical Cess model for closed-channel flow and pipe flow. These findings underscore the role of outer boundary conditions in shaping the outer-region eddy viscosity and, consequently, mean-flow predictions.
}


\setlength{\textheight}{23.6cm}
\thispagestyle{empty}

\maketitle
\setlength{\parindent}{1em}

\vspace{-1mm}
\begin{multicols}{2}

\section{Introduction}

Pressure-driven wall-bounded turbulence, sustained by an imposed streamwise pressure gradient (or an equivalent body force in numerical simulations), is ubiquitous in both natural and engineered systems and plays a central role in drag reduction and flow control \cite{xu2013coherent,procaccia2008colloquium}, as well as in mixing and transport processes \cite{zhang2024structure,tian2022lagrangian,huang2025how,chen2023backflow}. 
In environmental flows, wall turbulence with a free surface is directly relevant to rivers, lakes, and oceanic currents \cite{nezu2005open}. 
In mechanical engineering applications, wall turbulence accounts for a substantial fraction of the energy dissipated during fluid transport through pipes and channels \cite{jimenez2012cascades,kim2011physics,canton2016on,yao2018sojf}. \Authorfootnote
Canonical fully developed pressure-driven configurations, including plane closed-channel flow with no-slip walls at both the top and bottom, open-channel flow with a free-slip surface, and pipe flow, share a common driving mechanism but differ in geometry and boundary conditions. 
The dominant control parameter in these systems is the friction Reynolds number $Re_{\tau}$, which characterizes the intensity of near-wall shear \cite{lee2015retau5200,pirozzoli2021onepoint,yao2022ocf,cheng2024progress}, whereas a key response parameter is the skin-friction coefficient $C_f$, which quantifies the wall-shear drag normalized by the bulk flow \cite{xia2021skin}.

In wall turbulence, the mean streamwise velocity profile is a primary observable \cite{reichardt1951vollstaendige,she2017symmetry,cantwell2019universal}.
Its variation with wall-normal distance reflects the fact that different physical mechanisms dominate at different heights.
In the viscous sublayer, the velocity increases linearly with distance from the wall \cite{pope2000turbulent}.
In the overlap region, a logarithmic law emerges from the approximately constant Reynolds shear stress together with a mixing-length argument, according to which the characteristic turbulent length scale grows linearly with the wall distance on a semi-logarithmic scale \cite{prandtl1925bericht,vonkarman1931,pope2000turbulent,Smits2011high}.
Farther from the wall, the decay of stress and the saturation of outer length scales motivate velocity-defect and wake descriptions \cite{baldwinlomax1978,coles1956wake,townsend1976}.
Importantly, these classical viewpoints can also be reinterpreted in terms of the implied eddy-viscosity distribution $\nu_t(y)$, thereby providing a direct link between velocity laws and closure modeling.

For wall turbulence, simple ``zero-equation'' constructions remain attractive because they provide explicit links among $\nu_t(y)$, the mean momentum balance, and composite velocity laws, making them convenient baseline models for both Reynolds-averaged Navier--Stokes (RANS) simulations \cite{shan2024modeling} and wall-modeled large-eddy simulations (LES) \cite{Ciofalo2022,guo2012les}.
For example, Liu \emph{et al.} \cite{liu2025tsswm} developed an accurate total-shear-stress-conserving wall model that incorporates an eddy-viscosity modification and delivers robust predictions of skin friction and low-order turbulence statistics.
Among such ``zero-equation'' constructions, the classical Cess profile combines near-wall damping \cite{vandriest1956} with an outer-region Reichardt-type form \cite{reichardt1951vollstaendige} and has long served as a convenient baseline model \cite{cess1958survey,reynolds1967stability,symon2021energy,cossu2022onset,Fan2024Eddy}, as summarized in Sect.~\ref{sec:section2}.
In resolvent analysis, Ying \emph{et al.} \cite{ying2024eddy} optimize an eddy-viscosity profile by adjusting the classical Cess profile as a function of $Re_{\tau}$ and disturbance scales, thereby significantly improving the agreement between resolvent modes and DNS results.
A similar Cess-type approach is adopted by Sun \cite{sun2023poiseuille}, who extends a Prandtl--van Driest mixing-length closure to account for the effects of both the bottom and top walls.
Nevertheless, modern high-$Re_{\tau}$ databases \cite{hoyas2006scaling,hoyas2008budgets,bernardini2014retau4000,lee2015retau5200,hoyas2022highretau,hoyas2022turbulence,yao2022ocf,pirozzoli2023loglaw,yao2023pipe,pirozzoli2021onepoint,pirozzoli2024streamwisevar} suggest that the outer-region behavior of $\nu_t$ is not fully universal among closed-channel flow, open-channel flow, and pipe flow, thereby motivating a configuration-aware refinement of the outer-region eddy-viscosity representation.

Motivated by recent advances in high-$Re$ direct numerical simulations \cite{hoyas2006scaling,hoyas2008budgets,bernardini2014retau4000,lee2015retau5200,hoyas2022highretau,hoyas2022turbulence,yao2022ocf,pirozzoli2023loglaw,yao2023pipe,pirozzoli2021onepoint,pirozzoli2024streamwisevar} and renewed interest in eddy viscosity for turbulence modeling and analysis, we revisit $\nu_t$ in pressure-driven wall turbulence from a comparative perspective across configurations.
At low to moderate Reynolds numbers, limited scale separation causes inner viscous effects to contaminate the outer layer.
Only at sufficiently high Reynolds numbers does the geometry-dependent outer-layer profile become evident, enabling the extraction of asymptotically robust, configuration-specific outer-correction parameters.
Using the mean velocity, Reynolds shear stress, and the exact mean momentum balance, we infer full-depth or full-radius distributions of $\nu_t(y)$ for closed-channel flow, open-channel flow, and pipe flow.
We show that the classical Cess model \cite{cess1958survey} performs poorly in the outer region and therefore propose a revised outer correction function to represent the outer-region eddy-viscosity distribution.
Embedding this outer component into a Cess-type framework yields a closed-form expression for $\nu_t(y)$ that is applicable over the full depth or radius, providing a useful baseline for future studies of wall modeling, stability, and resolvent-based linear analysis across configurations.
The remainder of the paper is organized as follows. Section~\ref{sec:section2} summarizes representative eddy-viscosity models in wall turbulence. Section~\ref{sec:section3} describes the DNS databases for closed-channel flow, open-channel flow, and pipe flow. Section~\ref{sec:section4} presents a comprehensive analysis of eddy-viscosity distributions, modeling, and validation. Section~\ref{sec:section5} summarizes the main findings of the present work.

\section{Overview of representative eddy-viscous models in wall turbulence}\label{sec:section2}

Eddy-viscosity closures formalize the link between the mean momentum balance and turbulent transport through the Boussinesq hypothesis \cite{pope2000turbulent}, whereby the Reynolds shear stress is modeled as
  \begin{equation}
    \label{eq:boussinesq_stress}
    -\overline{u^\prime v^\prime} = \nu_t \frac{d\overline{u}}{dy},
  \end{equation}
where $u$ and $v$ denote the streamwise and wall-normal (or radial) velocity components, respectively, the overbar denotes time averaging, and primes denote fluctuations. A convenient interpretation is to express the eddy viscosity as the product of a velocity scale $u_c$ and a characteristic length scale $\ell$ \cite{pope2000turbulent}:
  \begin{equation}
    \label{eq:nu_t_scale_decomp}
    \nu_t = u_c\,\ell .
  \end{equation}
In the classical overlap region at sufficiently high Reynolds number, the viscous stress is negligible compared with the turbulent stress, i.e., $\nu\, d\overline{u}/dy \ll -\overline{u^\prime v^\prime}$, where $\nu$ is the kinematic viscosity.
The Reynolds shear stress is then approximately constant, $-\overline{u^\prime v^\prime} \approx u_\tau^2$, where $u_\tau$ is the friction velocity.
The velocity scale can then be taken as $u_c \approx \sqrt{-\overline{u^\prime v^\prime}} \approx u_\tau$.
Taking the mixing length to be $\ell=\ell_m=\kappa y$ \cite{prandtl1925bericht,vonkarman1931}, where $\kappa$ is the von K\'arm\'an constant, we obtain the classical log-law expressions for the eddy viscosity and mean shear:
  \begin{equation}
    \label{eq:loglaw_nut_and_shear}
    \nu_t = u_\tau \kappa y,
    \qquad
    \frac{d\overline{u}}{dy} = \frac{u_\tau}{\kappa y},
  \end{equation}
along with the log-law velocity profile:
  \begin{equation}\label{eq:loglaw}
    \overline{u}^+=(1/\kappa)\ln y^+ + B,
  \end{equation}
where $B$ is a constant; specifically, $B=5.2$ for closed-channel flow \cite{pope2000turbulent}, $B=4.21$ for open-channel flow \cite{pirozzoli2023loglaw}, and $B=4.53$ for pipe flow \cite{pirozzoli2021onepoint}.

However, the log-law scaling breaks down near the wall, where viscous effects become significant.
A widely used remedy is the van Driest damping function $g(y^+)$ \cite{vandriest1956}, which attenuates the effective length scale $\ell$ as $y^+ \to 0$ while recovering the overlap-region form $\ell_m=\kappa y$:
  \begin{equation}
    \label{eq:van_driest_damping}
    \ell(y)=\kappa y\, g(y^+), \qquad
    g(y^+)=1-e^{-y^+/A},
  \end{equation}
where $y^+=yu_\tau/\nu$ and $A$ is an empirical constant. Accordingly, the near-wall eddy viscosity may be written as
  \begin{equation}
    \label{eq:nearwall_nut_vandriest}
    \nu_t = u_\tau \kappa y\, g(y^+),
  \end{equation}
which reduces to Eq.~\eqref{eq:loglaw_nut_and_shear} for large $y^+$ and provides the correct damping as the wall is approached.

Outside the overlap region, the mean profile is influenced by outer-layer dynamics and geometry-specific boundary conditions, such as centerline symmetry in closed-channel and pipe flows and a free-slip surface in open-channel flow.
Reichardt \cite{reichardt1951vollstaendige} proposed a widely used composite mean-velocity representation for pipe flow that provides a smooth transition from the inner to the outer region, which may be written as
  \begin{equation}
    \label{eq:reichardt_nut_pipe}
    \frac{\nu_t}{\nu}=\frac{\kappa Re_\tau}{3}\,(0.5+r^2)(1-r^2),
  \end{equation}
where $r$ denotes the radial coordinate normalized by the pipe radius, and $Re_\tau=u_\tau \delta/\nu$ is the friction Reynolds number, with $\delta$ equal to the half-height $h$ for closed-channel flows, the full depth $h$ for open-channel flows, and the radius $R$ for pipe flow.
Although this representation was originally derived for pipe flow, incorporating it into the classical Cess model (see Eq.~\eqref{eq:cess_model}) allows it to be extended directly to closed-channel flow \cite{symon2021energy}, yielding a level of agreement with DNS similar to that obtained in pipe flow.

A complementary viewpoint is offered by Townsend's velocity-defect framework \cite{townsend1976}, which may be rationalized by assuming an approximately uniform eddy viscosity in the outer region:
  \begin{equation}
    \label{eq:townsend_outer_constant_nut}
    \nu_t \approx \nu_{t,c}=\mathrm{const.}
  \end{equation}
  In the outer region of fully developed pressure-driven wall flows, the Reynolds shear stress satisfies $-\overline{u^\prime v^\prime} \approx u_\tau^2(1-y/\delta)$. Substituting this relation into Eq.~\eqref{eq:boussinesq_stress} gives
  \begin{equation}
    \label{eq:outer_stress_balance}
    -\overline{u^\prime v^\prime}=\nu_t\frac{d\overline{u}}{dy}\approx u_\tau^2\left(1-\frac{y}{\delta}\right),
    \qquad
    \frac{d\overline{u}}{dy}\approx \frac{u_\tau^2}{\nu_{t,c}}\left(1-\frac{y}{\delta}\right).
  \end{equation}
  Introducing $\overline{u}^+ = \overline{u}/u_\tau$ then leads to
  \begin{equation}
    \label{eq:outer_uplus_gradient}
    \frac{d\overline{u}^+}{d(y/\delta)} \approx R_s\left(1-\frac{y}{\delta}\right),
    \qquad
    R_s=\frac{u_\tau\delta}{\nu_{t,c}},
  \end{equation}
  and integrating this expression from $y$ to the centerline at $y=\delta$ yields a quadratic defect form,
  \begin{equation}
    \label{eq:quadratic_defect}
    u_{\mathrm{CL}}^+ - u^+(y) \approx \frac{R_s}{2}\left(1-\frac{y}{\delta}\right)^2,
  \end{equation}
  where $u_{\mathrm{CL}}^+$ denotes the centerline velocity at $y^+=Re_\tau$. The prefactor $R_s$ is directly tied to the assumed outer eddy viscosity: a larger (more diffusive) $\nu_{t,c}$ implies a smaller defect amplitude.
  Empirically, $R_s$ takes values of 14 in closed-channel flow \cite{bernardini2014retau4000}, 16 in open-channel flow (based on a fit over $y/\delta \ge 0.6$), and 16 in pipe flow \cite{pirozzoli2021onepoint}.

\begin{figure*}[t]
\centering
\includegraphics[scale=0.8]{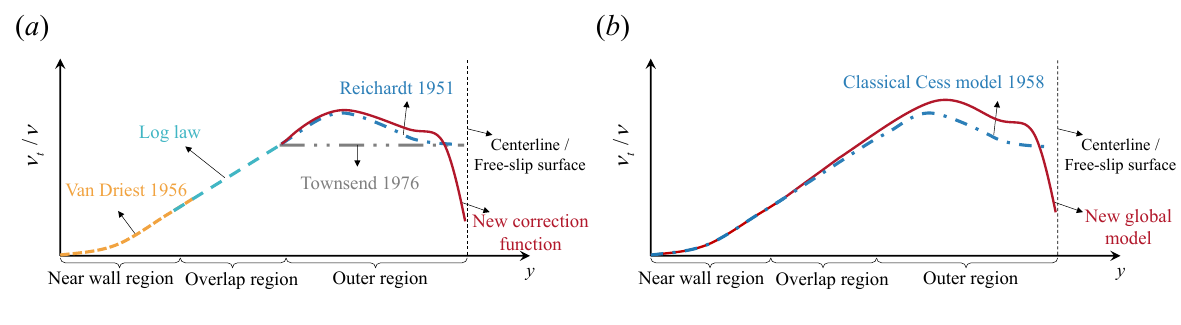}
    \caption{Schematic overview of representative eddy-viscosity models. ($a$) Region-wise constructions, in which different expressions are adopted in the near-wall, overlap, and outer regions. ($b$) Full-depth formulations obtained by combining near-wall and outer components. The classical Cess model (Eq.~\eqref{eq:cess_model}) combines a Reichardt-type outer representation (Eq.~\eqref{eq:reichardt_nut_pipe}) with a van Driest damping function in the near-wall region (Eq.~\eqref{eq:nearwall_nut_vandriest}). The present model retains the same near-wall damping (Eq.~\eqref{eq:nearwall_nut_vandriest}) but replaces the Cess/Reichardt outer component with the proposed outer correction function (Eq.~\eqref{eq:parametric_kappa_f} in Sect.~\ref{sec:section4.2}), thereby preserving identical near-wall behavior while introducing differences primarily in the outer region. The vertical dashed line indicates the centerline in closed-channel and pipe flows or the free-slip surface in open-channel flow.}
\label{fig:eddymodel}       
\end{figure*}

Full-depth algebraic eddy-viscosity models can be constructed by combining a near-wall damping component with an outer-layer representation in a single closed-form expression. A classical example is the Cess model \cite{cess1958survey,reynolds1967stability}, which blends van Driest-type damping (Eq.~\eqref{eq:nearwall_nut_vandriest}) with a Reichardt-type outer structure (Eq.~\eqref{eq:reichardt_nut_pipe}) and may be written compactly as  
\begin{equation}
\frac{\nu_t}{\nu} = \frac{1}{2} \left\{ 1 + \left[ \frac{\kappa Re_\tau}{3} (1 - z_c^2)(1 + 2z_c^2) \left(1 - e^{-z_c^+ / A}\right) \right]^2 \right\}^{1/2} - \frac{1}{2},
\label{eq:cess_model}
\end{equation}  
where $z_c=1-y/\delta$, $z_c^+=Re_\tau\left(1-\left|y/\delta\right|\right)$, and $A$ is the same damping constant as in Eq.~\eqref{eq:nearwall_nut_vandriest}. In the present study, fitting across datasets for $y^+<100$ gives $A=26.95$ for closed-channel flow, $A=24.88$ for open-channel flow, and $A=25.71$ for pipe flow.
Figure~\ref{fig:eddymodel} summarizes these representative constructions in the near-wall, overlap, and outer regions and highlights how region-wise building blocks can be combined into full-depth expressions, thereby providing a baseline for subsequent modifications targeting outer-region behavior under different boundary conditions.

  Beyond algebraic models, eddy viscosity can also be obtained from transport-equation closures. A prototypical example is the two-equation $k$--$\varepsilon$ model \cite{launder1983chapter}, for which
  \begin{equation}
    \label{eq:kepsilon_nut}
    \nu_t = C_\mu \frac{k^2}{\varepsilon},
  \end{equation}
  where $k$ is the turbulent kinetic energy, $\varepsilon$ is the dissipation rate, and $C_\mu \approx 0.09$ in the standard formulation. For the datasets considered here, independent fits to the closed-channel and pipe-flow cases yield $C_\mu = 0.06$ for both configurations. This transport-based baseline is included for comparison with the algebraic eddy-viscosity representations introduced above.

\section{DNS databases of pressure-driven wall turbulence at high-Reynolds-number}\label{sec:section3}

This study draws on publicly available direct numerical simulation (DNS) databases for three canonical pressure-driven wall-bounded flows: plane closed-channel flow, open-channel flow with a free-slip surface, and pipe flow.
The selected datasets span friction Reynolds numbers from $Re_{\tau} \approx 2000$ to $12000$ and provide one-point statistics, including the mean velocity and Reynolds shear stress. In some cases, higher-order statistics and Reynolds-stress budgets are also available and are used for cross-checks and model diagnostics.
For clarity of presentation and to facilitate inter-configuration comparisons, we group the cases into seven nominal friction-Reynolds-number bins: $Re_{\tau} \approx 2000, 3000, 4000, 5000, 6000, 10000,$ and $12000$.

  \subsection{Key definitions and notation}
For all configurations, we adopt inner scaling based on the friction velocity:
  \begin{equation}
    \label{eq:friction_velocity}
    u_\tau = \sqrt{\tau_w/\rho},
  \end{equation}
where $\tau_w$ is the mean wall shear stress, and $\rho$ is the fluid density. The friction Reynolds number is defined as
  \begin{equation}
    \label{eq:ret_definition}
    Re_\tau = \frac{u_\tau \delta}{\nu},
  \end{equation}
where $\delta$ denotes the maximum distance from the wall, namely the half-height $h$ for closed-channel flow, the full depth $h$ for open-channel flow, and the radius $R$ for pipe flow. Inner (``$+$'') variables are normalized by the viscous length scale $\delta_\nu=\nu/u_\tau$. Accordingly, $u^+=u/u_\tau$ and $y^+=y/\delta_\nu$, where $y$ denotes the distance from the nearest solid wall.
For closed-channel and open-channel flows, we use Cartesian coordinates $(x,y,z)$ for the streamwise, wall-normal, and spanwise directions, respectively. For pipe flow, we use cylindrical coordinates $(x,r,\theta)$, where $x$ is the axial (streamwise) direction, $r$ is the radial coordinate measured from the centerline, and $\theta$ is the azimuthal direction. The wall-normal distance is then defined as $y=R-r$, so that $y^+=(R-r)/\delta_\nu$.

  \subsection{Closed-channel flow databases}
  For closed-channel flow, we draw on four complementary DNS datasets: (1) Hoyas and Jim\'enez \cite{hoyas2006scaling,hoyas2008budgets} (up to $Re_{\tau}\approx 2003$); (2) Bernardini \emph{et al.} \cite{bernardini2014retau4000} (up to $Re_{\tau}\approx 4079$); (3) Lee and Moser \cite{lee2015retau5200} (up to $Re_{\tau}\approx 5186$); (4) Hoyas \emph{et al.} \cite{hoyas2022highretau} together with Oberlack \emph{et al.} \cite{hoyas2022turbulence} (up to $Re_{\tau}\approx 10000$). The key simulation parameters are summarized in Table~\ref{tab:ccf_dns_datasets}.

\begin{table*}[t]
\centering
\small
\setlength{\tabcolsep}{3pt}
\renewcommand{\arraystretch}{0.95}

    \caption{Summary of the closed-channel-flow DNS datasets used in this study. The table lists the case identifier, friction Reynolds number $Re_{\tau}$, computational domain size $L_x \times L_y \times L_z$, streamwise and spanwise grid resolutions in inner units ($\Delta x^+$ and $\Delta z^+$), the wall-normal grid spacing at the wall and its maximum value ($\Delta y_w^+$ and $\Delta y_{\text{max}}^+$), the number of wall-normal grid points $N_y$, and the eddy-turnover time (ETT) covered by the statistically sampled record. The last two columns give the reference source and the plotting symbol adopted throughout the paper. This table follows the same format as Table~\ref{tab:ocf_dns_datasets}.}
    \label{tab:ccf_dns_datasets}

\begin{tabular}{c c c c c c c c c c c}
\hline
Case & $Re_{\tau}$ & $L_x \times L_y \times L_z$ & $\Delta x^+$ & $\Delta z^+$ &
$\Delta y_w^+$ & $\Delta  y_{\text{max}}^+$ & $N_y$ & ETT & Ref & Symbol \\
\hline

CCF1Re2000  & 2003 & $8\pi h\times 2h\times 3\pi h$ & 12  & 6.1 & --   & 8.9 & 633  & 11    & Hoyas and Jim\'enez \cite{hoyas2006scaling,hoyas2008budgets} &
$\square$ \\
\hline

CCF2Re2000  & 2022 & $6\pi h\times 2h\times 2\pi h$ & 9.3  & 6.2 & 0.01 & --  & 768  & 14.9 &
\multirow{2}{*}{Bernardini \emph{et al.} \cite{bernardini2014retau4000}} &
\multirow{2}{*}{$\bigcirc$} \\
CCF2Re4000  & 4079 & $6\pi h\times 2h\times 2\pi h$ & 9.4  & 6.2 & 0.01 & --  & 1024 & 8.54 & & \\
\hline

CCF3Re5000  & 5186 & $8\pi h\times 2h\times 3\pi h$ & 12.7 & 6.4 & 0.498& 10.3& 1536 & 7.80 & Lee and Moser \cite{lee2015retau5200} & $\triangle$ \\
\hline

CCF4Re10000 & 10000& $2\pi h\times 2h\times \pi h$   & 15.3 & 7.6 & 0.3  & 12  & 2101 & 19.8 &
Hoyas \emph{et al.} \cite{hoyas2022highretau}, Oberlack \emph{et al.} \cite{hoyas2022turbulence} & \ostarsix \\
\hline
\end{tabular}
\end{table*}

  \subsection{Open-channel flow databases}
  For open-channel flow, we draw on two complementary DNS datasets: (1) Yao \emph{et al.} \cite{yao2022ocf} (up to $Re_{\tau}\approx 2006$) and (2) Pirozzoli \cite{pirozzoli2023loglaw} (up to $Re_{\tau}\approx 6010$). The key simulation parameters are summarized in Table~\ref{tab:ocf_dns_datasets}.

\begin{table*}[t]
\centering
\small
\setlength{\tabcolsep}{3pt}
\renewcommand{\arraystretch}{0.95}

    \caption{Summary of the open-channel-flow DNS datasets used in this study.}
    \label{tab:ocf_dns_datasets}

\begin{tabular}{c c c c c c c c c c c}
\hline
Case & $Re_{\tau}$ & $L_x \times L_y \times L_z$ & $\Delta x^+$ & $\Delta z^+$ &
$\Delta y_w^+$ & $\Delta y_{\max}^+$ & $N_y$ & ETT & Ref & Symbol \\
\hline

OCF1Re2000 & 2006 & $8\pi h\times h\times 4\pi h$ & 12.3 & 6.2 & 0.09 & 3.8 & 768 & 10.1 & Yao \emph{et al.} \cite{yao2022ocf} & $\bigtriangledown$ \\
\hline

OCF2Re2000 & 2002 & $6\pi h\times h\times 2\pi h$ & 8.5 & 4.0 & $<0.1$ & -- & 265 & 27.6 &
\multirow{3}{*}{Pirozzoli \cite{pirozzoli2023loglaw}} &
\multirow{3}{*}{$\lozenge$} \\
OCF2Re3000 & 3009 & $6\pi h\times h\times 2\pi h$ & 8.5 & 4.0 & $<0.1$ & -- & 355 & 44.0 & & \\
OCF2Re6000 & 6010 & $6\pi h\times h\times 2\pi h$ & 8.5 & 4.0 & $<0.1$ & -- & 591 & 23.4 & & \\
\hline

\end{tabular}
\end{table*}

  \subsection{Pipe flow databases}
  For pipe flow, we draw on three complementary DNS datasets: (1) Yao \emph{et al.} \cite{yao2023pipe} (up to $Re_{\tau}\approx 5197$), (2) Pirozzoli \emph{et al.} \cite{pirozzoli2021onepoint} (up to $Re_{\tau}\approx 6019$), and (3) Pirozzoli \cite{pirozzoli2024streamwisevar} (up to $Re_{\tau}\approx 12055$). The key simulation parameters are summarized in Table~\ref{tab:pf_dns_datasets}.

\begin{table*}[t]
\centering
\small
\setlength{\tabcolsep}{3pt}
\renewcommand{\arraystretch}{0.95}

    \caption{Summary of the pipe-flow DNS datasets used in this study. The table lists the case identifier, friction Reynolds number $Re_{\tau}$, axial length $L_x$, axial and azimuthal grid resolutions in inner units ($\Delta x^+$ and $\Delta (R\theta)^+$), the radial grid spacing at the wall and its maximum value ($\Delta r_w^+$ and $\Delta r_{\text{max}}^+$), the number of radial grid points $N_r$, and the eddy-turnover time (ETT) covered by the statistically sampled record. The last two columns give the reference source and the plotting symbol adopted throughout the paper.}
    \label{tab:pf_dns_datasets}

\begin{tabular}{c c c c c c c c c c c}
\hline
Case & $Re_{\tau}$ & $L_x$ & $\Delta x^+$ & $\Delta (R\theta)^+$ &
$\Delta r_w^+$ & $\Delta r_{\text{max}}^+$ & $N_r$ & ETT & Ref & Symbol \\
\hline

PF1Re2000 & 2001 & $10\pi R$ & 10.2 & 4.9 & 0.1 & 3.9 & 768  & 9.7 &
\multirow{2}{*}{Yao \emph{et al.} \cite{yao2023pipe}} &
\multirow{2}{*}{\ostarfive} \\
PF1Re5000 & 5197 & $10\pi R$ & 12.8 & 6.3 & 0.2 & 8.6 & 1024 & 4.6  & & \\
\hline

PF2Re2000 & 1976 & $15R$ & 10 & 4.5 & 0.05 & -- & 399 & 22.4 &
\multirow{3}{*}{Pirozzoli \emph{et al.} \cite{pirozzoli2021onepoint}} &
\multirow{3}{*}{\raisebox{0.1ex}{\scalebox{1.35}{$\triangleright$}}} \\
PF2Re3000 & 3028 & $15R$ & 10 & 4.5 & 0.05 & -- & 540 & 16.6 & & \\
PF2Re6000 & 6019 & $15R$ & 10 & 4.5 & 0.05 & -- & 910 & 8.32 & & \\
\hline

PF3Re2000  & 1972  & $15R$ & 10 & 4.1 & 0.05 & -- & 243  & 45.1 &
\multirow{4}{*}{Pirozzoli \cite{pirozzoli2024streamwisevar}} &
\multirow{4}{*}{\raisebox{0.1ex}{\scalebox{1.35}{$\triangleleft$}}} \\
PF3Re3000  & 3027  & $15R$ & 10 & 4.1 & 0.05 & -- & 327  & 26.9 & & \\
PF3Re6000  & 6006  & $15R$ & 10 & 4.1 & 0.05 & -- & 546  & 18.2 & & \\
PF3Re12000 & 12055 & $15R$ & 10 & 4.1 & 0.05 & -- & 1024 & 6.99 & & \\
\hline

\end{tabular}
\end{table*}

\section{Results and discussion}\label{sec:section4}
\subsection{Wall-normal profile of eddy viscosity}\label{sec:section4.1}

\begin{figure*}[t]
\centering
\includegraphics[scale=0.8]{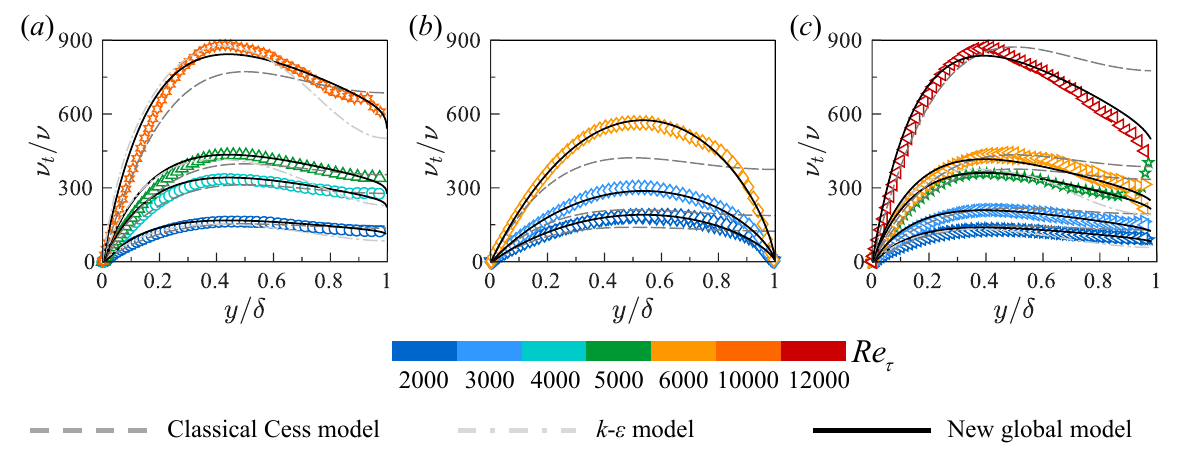}
    \caption{Wall-normal profiles of the eddy viscosity, $\nu_t/\nu$, for (a) plane closed-channel flow, (b) open-channel flow, and (c) pipe flow at representative values of $Re_{\tau}$. The results are compared with the classical Cess model (Eq.~\eqref{eq:cess_model}), a DNS-based $k$--$\varepsilon$ estimate (Eq.~\eqref{eq:kepsilon_nut}), and the new global model (Eq.~\eqref{eq:new_global_model}), which is formulated within a Cess-type framework using the proposed outer correction function (Eq.~\eqref{eq:parametric_kappa_f}). Symbols denote DNS data from Tables~\ref{tab:ccf_dns_datasets}-\ref{tab:pf_dns_datasets}. The wall-normal coordinate is normalized by $\delta$, where $\delta=h$ for closed-channel and open-channel flows and $\delta=R$ for pipe flow; in the latter case, $y=R-r$.}
\label{fig:eddymodelDNSCompares}       
\end{figure*}

Figure~\ref{fig:eddymodelDNSCompares} compares the wall-normal profiles of eddy viscosity in pressure-driven wall turbulence for closed-channel flow, open-channel flow, and pipe flow.
Using DNS one-point statistics, an effective eddy viscosity can be inferred from Eq.~\eqref{eq:boussinesq_stress} as $\nu_t(y)=-\overline{u^\prime v^\prime}/(d\overline{u}/dy)$ and is reported in nondimensional form as $\nu_t/\nu$ versus the outer coordinate $y/\delta$.
At the outer boundary, $y=\delta$, the mean velocity gradient vanishes, leading to a formal $0/0$ singularity in the Boussinesq relation.
Consequently, the DNS-inferred eddy viscosity evaluated at the centerline or free surface is sensitive to numerical differentiation errors and often exhibits unphysical jumps or scatter, particularly in pipe flow owing to the cylindrical grid arrangement.
Therefore, although the physical limiting behavior is a finite $\nu_t$ under centerline symmetry and a vanishing $\nu_t$ at a free-slip surface, data points in the immediate vicinity of the boundary are regarded as numerically unreliable.
In all subsequent quantitative fitting procedures, we exclude the final grid point for closed- and open-channel flows and impose an upper cutoff of $y/\delta=0.98$ for pipe flow to ensure robust calibration.

As shown in Fig.~\ref{fig:eddymodelDNSCompares}, $\nu_t/\nu$ increases rapidly away from the wall in all three configurations and reaches a maximum of order $10^2$--$10^3$ in viscous units, reflecting the dominance of turbulent transport over most of the cross-section at high Reynolds numbers.
However, the outer-region behavior is not universal across the three configurations. In particular, both the wall-normal variation and the magnitude of $\nu_t/\nu$ near the outer boundary (i.e., the centerline in closed-channel and pipe flows, and the free surface in open-channel flow) depend on the configuration, indicating that Townsend's assumption of a uniform outer eddy viscosity \cite{townsend1976} is not generally valid.
Moreover, neither the classical Cess model (Eq.~\eqref{eq:cess_model}) nor a DNS-based $k$--$\varepsilon$ estimate (Eq.~\eqref{eq:kepsilon_nut}) reproduces the DNS-inferred eddy-viscosity distribution consistently across all three configurations, with the largest discrepancies occurring in the outer region.
This limitation of the classical Cess model (Eq.~\eqref{eq:cess_model}) is unsurprising, because the Reichardt outer representation (Eq.~\eqref{eq:reichardt_nut_pipe}) was originally developed for pipe flow and calibrated at comparatively low Reynolds numbers.
Motivated by these observations, we introduce a DNS-informed outer correction function (Sect.~\ref{sec:section4.2}, Eq.~\eqref{eq:parametric_kappa_f}) that captures the configuration-dependent outer behavior at high Reynolds numbers and, when embedded in a Cess-type framework (Sect.~\ref{sec:section4.2}, Eq.~\eqref{eq:new_global_model}), yields improved full-depth predictions for closed-channel flow, open-channel flow, and pipe flow.

\subsection{Outer correction function and a new global model}\label{sec:section4.2}

\begin{figure*}[t]
\centering
\includegraphics[scale=0.8]{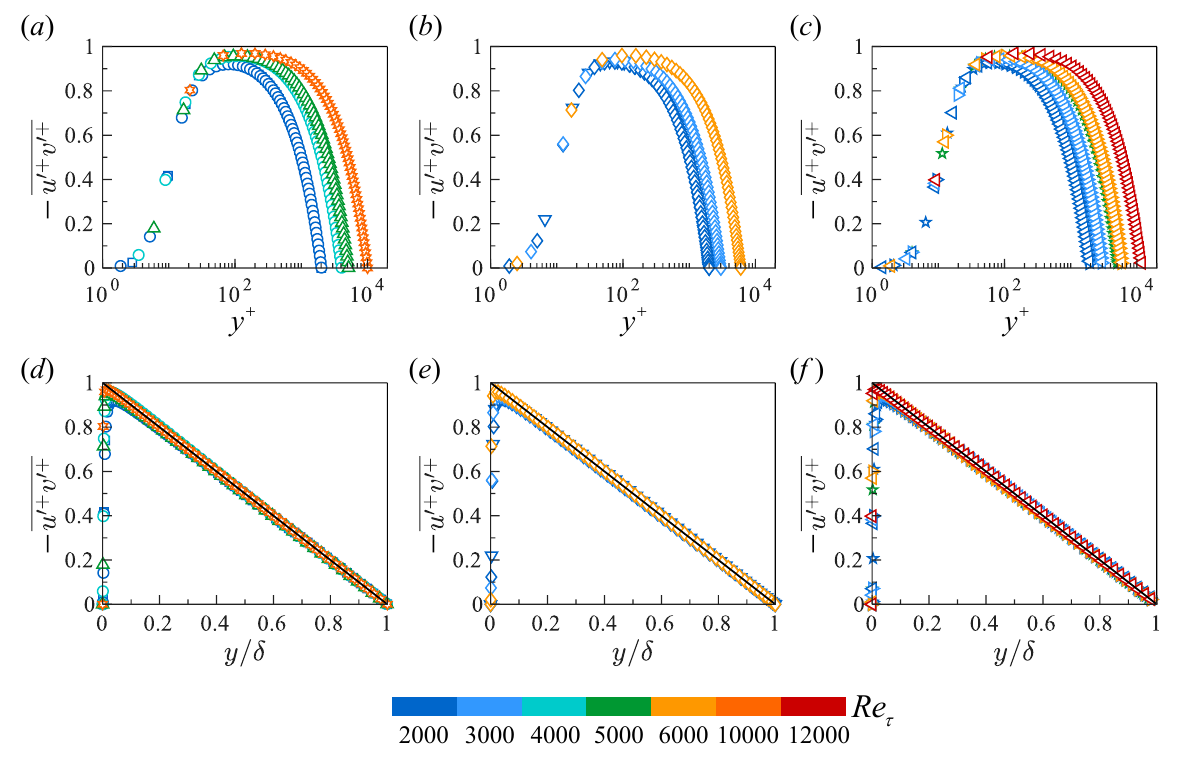}
    \caption{Wall-normal profiles of the Reynolds shear stress, $-\overline{u^{\prime +}v^{\prime +}}$, for ((a), (d)) closed-channel flow, ((b), (e)) open-channel flow, and ((c), (f)) pipe flow at representative $Re_{\tau}$. The DNS data are shown in semi-logarithmic coordinates in panels (a)-(c) to highlight the near-constant-stress layer and in linear coordinates in panels (d)-(f) to emphasize the outer-region variation. Symbols denote DNS results from Tables~\ref{tab:ccf_dns_datasets}-\ref{tab:pf_dns_datasets}. The black solid line in panels (d)-(f) denotes the exact total-stress distribution, $1-y/\delta$, where $\delta=h$ for closed- and open-channel flows and $\delta=R$ for pipe flow, with $y=R-r$.}
\label{fig:ReynoldsStress}       
\end{figure*}

To develop an eddy-viscosity model that is robust across configurations, we first examine the wall-normal distribution of the Reynolds shear stress.
For incompressible, fully developed pressure-driven flows, the mean streamwise momentum balance yields an exact expression for the total shear stress, i.e., the sum of the viscous and Reynolds stresses:
  \begin{equation}
    \label{eq:total_shear_stress_def}
    \tau_{\mathrm{tot}}(y)=\rho\left(\nu\,\frac{d\overline{u}}{dy}-\overline{u^\prime v^\prime}\right),
  \end{equation}
which varies linearly with wall-normal distance. Specifically, for pressure-driven wall turbulence, $\tau_{\mathrm{tot}}(y)=\rho u_\tau^2(1-y/\delta)$; normalizing by the wall shear stress $\tau_w=\rho u_\tau^2$ then gives
  \begin{equation}
    \label{eq:total_shear_stress_linear}
    \frac{\tau_{\mathrm{tot}}}{\tau_w}=1-\frac{y}{\delta}.
  \end{equation}
Figure~\ref{fig:ReynoldsStress} shows the Reynolds shear stress, $-\overline{u^{\prime +}v^{\prime +}}$, extracted from DNS.
In semi-logarithmic coordinates (Fig.~\ref{fig:ReynoldsStress}(a)-(c)), the Reynolds shear stress exhibits a broad, nearly constant plateau over an intermediate range, typically beginning around $y^+\approx 100$ at sufficiently high Reynolds numbers \cite{lee2015retau5200,pirozzoli2021onepoint,cui2025electric}.
The behavior of the peak Reynolds stress may also be interpreted in light of the work of Chen and Sreenivasan \cite{chen2021peak,chen2022bounded,chen2023asymptotics,chen2025pnas}, who proposed a unified asymptotic interpretation of wall-turbulence fluctuations in which near-wall dissipation, constrained by an upper bound on near-wall production in wall units, remains finite as the Reynolds number increases.
Within this framework, several canonical near-wall statistics, including peak turbulence intensities, approach finite high-$Re$ limits, with finite-$Re$ effects following a consistent defect-type scaling. Using this scaling, they rationalized complete wall-normal profiles and, more recently, extended the analysis to higher-order velocity moments. This framework therefore provides a reliable alternative to generalized logarithmic representations based on attached-eddy ideas \cite{townsend1976} for channels, pipes, and turbulent boundary layers.
In linear coordinates (Fig.~\ref{fig:ReynoldsStress}(d)-(f)), as $y/\delta$ increases, the viscous stress $\rho \nu\, d\overline{u}/dy$ becomes negligible, and the Reynolds stress $-\rho\overline{u^\prime v^\prime}$ closely follows the total shear stress, which decreases approximately linearly \cite{townsend1976}.

Following the velocity- and length-scale interpretation underlying the log law \cite{pope2000turbulent}, we extend the eddy-viscosity representation into the outer region.
Specifically, we modify the velocity scale to reflect the approximately linear stress distribution in the outer layer.
Because $-\rho\overline{u^\prime v^\prime}\approx \tau_{\mathrm{tot}}=\tau_w(1-y/\delta)$ in the outer region, we define a local velocity scale as
  \begin{equation}
    \label{eq:outer_velocity_scale_uc}
    u_c(y)\approx \sqrt{-\overline{u^\prime v^\prime}}\approx u_\tau\sqrt{1-\frac{y}{\delta}}.
  \end{equation}
We retain the mixing-length form for the characteristic length scale, $\ell_m=\kappa y$, because the wall distance remains the relevant geometric measure governing shear-driven mixing over much of the cross-section. However, the DNS data in Fig.~\ref{fig:ReynoldsStress} indicate that the outer behavior depends primarily on the outer coordinate $y/\delta$.
In the logarithmic layer, by contrast, the mixing-length argument gives $\nu_t/\nu=(u_\tau \kappa y)/\nu=\kappa y^+$ (Eq.~\eqref{eq:loglaw_nut_and_shear}), implying that the $\ell_m=\kappa y$ scaling alone cannot capture the configuration-dependent variations observed in the outer region.
We therefore introduce an outer correction factor $f(y/\delta)$ to account for outer-scaling effects in the turbulent length scale. The outer-region eddy viscosity is then written as the product of a velocity scale $u_c(y)$ and a length scale $\ell(y)$ \cite{prandtl1925bericht,pope2000turbulent}:
  \begin{equation}
    \begin{split}\label{eq:outer_nut_with_correction}
      \nu_t &= u_c(y)\,\ell(y) \\
            &= u_c(y)\,\ell_m\,f\!\left(\frac{y}{\delta}\right) \\
            &= \kappa u_\tau y\,\sqrt{1-\frac{y}{\delta}}\; f\!\left(\frac{y}{\delta}\right).
    \end{split}
  \end{equation}
  This relation provides a convenient diagnostic for extracting the correction function directly from DNS one-point statistics:
  \begin{equation}
    \label{eq:diagnostic_kappa_f}
    \kappa f\!\left(\frac{y}{\delta}\right)=\frac{\nu_t/\nu}{y^+\sqrt{1-y/\delta}},
  \end{equation}
  where $\nu_t(y)$ is inferred from DNS as $\nu_t=-\overline{u^\prime v^\prime}/(d\overline{u}/dy)$.

\begin{figure*}[t]
\centering
\includegraphics[scale=0.8]{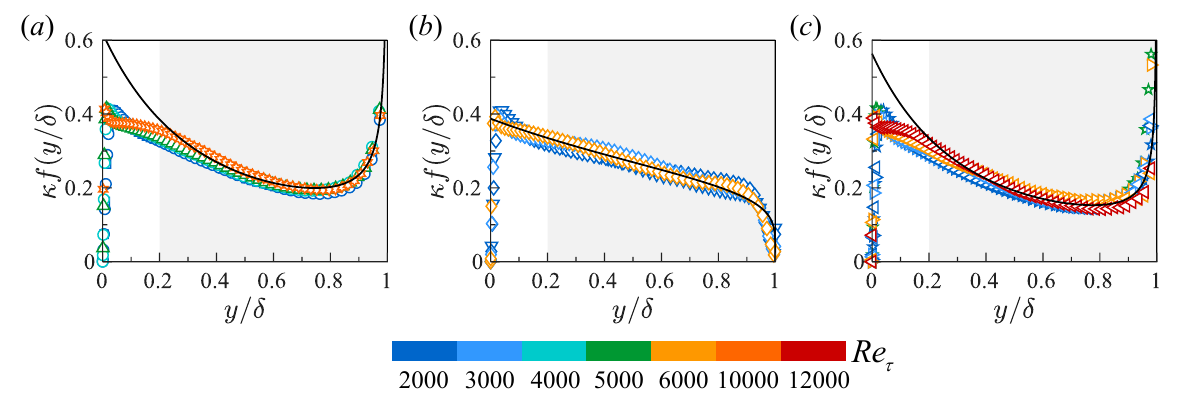}
    \caption{Wall-normal profiles of the scaled outer correction function, $\kappa f(y/\delta)$, for (a) closed-channel flow, (b) open-channel flow, and (c) pipe flow at representative $Re_{\tau}$. Symbols denote DNS results from Tables~\ref{tab:ccf_dns_datasets}-\ref{tab:pf_dns_datasets}. The black solid line denotes the fit $\kappa f(y/\delta)=\kappa\alpha(1-y/\delta)^p(1+y/\delta)^q$. The shaded region indicates the fitting interval $0.20 \le y/\delta < 1$, where $\delta=h$ for closed- and open-channel flows and $\delta=R$ for pipe flow, with $y=R-r$.}
\label{fig:correctionFunction}       
\end{figure*}

Figure~\ref{fig:correctionFunction} shows the resulting profiles of $\kappa f(y/\delta)$ for closed-channel flow, open-channel flow, and pipe flow. Within the outer region (i.e., the shaded region in Fig.~\ref{fig:correctionFunction}), the extracted $\kappa f(y/\delta)$ exhibits only weak Reynolds-number dependence, whereas its wall-normal variation is clearly configuration dependent.
In closed-channel flow and pipe flow, $\kappa f(y/\delta)$ decreases with increasing $y/\delta$ over a substantial portion of the cross-section and then turns upward as the centerline is approached (see Fig.~\ref{fig:correctionFunction}(a) and (c)), indicating a non-monotonic trend.
By contrast, in open-channel flow, $\kappa f(y/\delta)$ decreases monotonically toward the free surface, reflecting the different outer boundary condition and the associated redistribution of turbulent transport.
These trends help explain why the Reichardt outer representation (Eq.~\eqref{eq:reichardt_nut_pipe}) does not generalize to pressure-driven wall flows without a configuration-specific correction.
To capture the observed behavior in a compact analytic form, we propose
  \begin{equation}
    \label{eq:parametric_kappa_f}
    \kappa f\!\left(\frac{y}{\delta}\right)=\kappa \alpha \left(1-\frac{y}{\delta}\right)^{p}\left(1+\frac{y}{\delta}\right)^{q},
  \end{equation}
where the parameters $\alpha$, $p$, and $q$ are calibrated against the DNS data.

It should be acknowledged that Eq.~\eqref{eq:parametric_kappa_f} is a phenomenological, data-informed \textit{ansatz}, rather than a result derived from first principles such as asymptotic theory or attached-eddy arguments.
Nevertheless, its construction is guided by physical constraints and by the historical development of eddy-viscosity modeling.
The proposed formulation follows the philosophy of the classical Cess framework, which likewise introduces a multiplicative outer correction through the phenomenological Reichardt polynomial $(1-z_c^2)(1+2z_c^2)$ (Eq.~\eqref{eq:reichardt_nut_pipe}).
However, this polynomial form is overly rigid because it is hardwired to enforce centerline symmetry and is therefore unable to accommodate the monotonic, free-surface-driven decay observed in open-channel flow.
By generalizing the outer-coordinate dependence into adjustable power-law factors, the present form provides a compact extension of the classical outer representation.
Specifically, the term $(1-y/\delta)^p$ acts as an asymptotic controller near the outer boundary, determining whether the eddy viscosity vanishes explicitly, as in open-channel flow, or remains quasi-finite, as in symmetric closed-channel and pipe flows; the term $(1+y/\delta)^q$ modulates the interior wake-like redistribution of turbulent scales.
Although this representation introduces more free parameters than the Reichardt model, these parameters can be treated as configuration-specific constants for a given flow type, rather than as Reynolds-number-dependent tuning parameters.
We do not claim that this formulation is mathematically unique; rather, we adopt it because it offers a favorable balance among analytical simplicity, physical interpretability, and the flexibility required to capture the diverse configuration-dependent outer-layer trends observed in the DNS data.

In practice, the parameters $\alpha$, $p$, and $q$ are determined by fitting the DNS data over the relevant outer-region interval, $0.20 \le y/\delta < 1$, excluding the singular region near the boundary discussed in Sect.~\ref{sec:section4.1}.
The fitted values are listed in Table~\ref{tab:outer_correction_params}.
The choice of this interval reflects a compromise between physical consistency and full-depth predictive accuracy.
As reported by Lee and Moser~\cite{lee2015retau5200}, the logarithmic region extends to $y^+\approx 800$ at $Re_\tau\approx 5200$, corresponding to $y/\delta\approx 0.15$.
Thus, setting the lower bound to $y/\delta=0.20$ safely excludes the logarithmic region and ensures that the fitting targets only outer-region behavior, for which the linear total-stress assumption is robust.
Sensitivity tests show that narrowing the interval further artificially increases the parameter $\alpha$ in closed-channel and pipe flows, leading to an overprediction of the reconstructed mean velocity.
The resulting fit over $0.20 \le y/\delta < 1$ reproduces the DNS-inferred $\kappa f(y/\delta)$ well across all three configurations and thus provides an essential ingredient for constructing the global eddy-viscosity model.

\begin{table*}[t]
\centering
\small
\setlength{\tabcolsep}{6pt}
\renewcommand{\arraystretch}{1.05}
\caption{Parameters used in the outer correction function (Eq.~\eqref{eq:parametric_kappa_f}) and the proposed global model (Eq.~\eqref{eq:new_global_model}). The von K\'arm\'an constant, $\kappa$, is taken from Ref.~\cite{pope2000turbulent} for closed-channel flow, from Ref.~\cite{pirozzoli2023loglaw} for open-channel flow, and from Ref.~\cite{pirozzoli2021onepoint} for pipe flow. The parameters $\alpha$, $p$, $q$, and the damping constant $A$ are calibrated over the fitting interval $0.20 \le y/\delta < 1$.}
\label{tab:outer_correction_params}
\begin{tabular}{lccccc}
\hline
System & $\kappa$ & $\alpha$ & $p$ & $q$ & $A$ \\
\hline
Closed-channel flow & 0.410 & 1.51 & $-0.46$ & $-3.17$ & 44.33 \\
Open-channel flow   & 0.375 & 1.03 & $\phantom{-}0.20$ & $-0.54$  & 26.06 \\
Pipe flow           & 0.387 & 1.46 & $-0.43$ & $-3.38$ & 40.46 \\
\hline
\end{tabular}
\end{table*}

  We now construct an explicit full-depth eddy-viscosity expression by combining a near-wall damping function with the present outer correction, following the classical Cess-type methodology~\cite{reynolds1967stability,sun2023poiseuille}. Using the total shear-stress relation (Eq.~\eqref{eq:total_shear_stress_def}) together with the Boussinesq hypothesis (Eq.~\eqref{eq:boussinesq_stress}), the mean momentum balance can be written in the compact form:
  \begin{equation}
    \label{eq:compact_momentum_balance}
    \left(\nu+\nu_t\right)\frac{d\overline{u}}{dy}=\frac{\tau_{\mathrm{tot}}}{\rho}.
  \end{equation}
  To obtain an analytical closure for $\nu_t$ while retaining full-depth applicability, we adopt a Prandtl-type mixing-length argument. Introducing a characteristic length $\ell(y)$, we define the associated velocity scale based on the local mean shear as
  \begin{equation}
    \label{eq:uc_def_shear}
    u_c=\ell\left|\frac{d\overline{u}}{dy}\right|.
  \end{equation}
  The same velocity scale may also be estimated from the Reynolds shear stress as $-\overline{u^\prime v^\prime}\approx u_c^2$, which gives
  \begin{equation}
    \label{eq:reynolds_stress_uc}
    -\overline{u^\prime v^\prime}=\ell^2\left(\frac{d\overline{u}}{dy}\right)^2.
  \end{equation}
  Combining this estimate with the Boussinesq relation yields a convenient identity linking the eddy viscosity to the characteristic length,
  \begin{equation}
    \label{eq:nut_ell_relation}
    \nu_t=\ell^2\frac{d\overline{u}}{dy}.
  \end{equation}
  Substituting Eq.~\eqref{eq:nut_ell_relation} into Eq.~\eqref{eq:compact_momentum_balance} yields a quadratic equation for the mean shear, $S\equiv d\overline{u}/dy$:
  \begin{equation}
    \label{eq:quadratic_for_S}
    \ell^2 S^2+\nu S-\frac{\tau_{\mathrm{tot}}}{\rho}=0.
  \end{equation}
  Because $\overline{u}$ increases away from the wall, the mean shear is positive, and we therefore take
  \begin{equation}
    \label{eq:S_solution}
    S=\frac{-\nu+\sqrt{\nu^2+4\ell^2\,\tau_{\mathrm{tot}}/\rho}}{2\ell^2}.
  \end{equation}
  The corresponding eddy viscosity then follows as
  \begin{equation}
    \label{eq:nut_solution_dimensional}
    \nu_t=\ell^2 S=\frac{-\nu+\sqrt{\nu^2+4\ell^2\,\tau_{\mathrm{tot}}/\rho}}{2}.
  \end{equation}
  Using Eq.~\eqref{eq:total_shear_stress_linear} together with the viscous length scale $\delta_\nu=\nu/u_\tau$, this expression can be written in nondimensional form as
  \begin{equation}
    \label{eq:nut_nondim_general}
    \frac{\nu_t}{\nu}
    =\frac{1}{2}\left[\sqrt{1+4(\ell^+)^2\left(1-\frac{y}{\delta}\right)}-1\right],
  \end{equation}
  where $\ell^+\equiv \ell/\delta_\nu=\ell u_\tau/\nu$.

  To complete the eddy-viscosity model, it remains to specify a continuous full-depth characteristic length scale $\ell(y)$. In the near-wall region, we adopt a van Driest-type damping function $g(y^+)$ given by Eq.~\eqref{eq:van_driest_damping} to recover the correct viscous- and buffer-layer behavior, while retaining the standard mixing-length scaling $\ell_m=\kappa y$ as the baseline length scale.
  To incorporate the configuration-dependent outer behavior identified in Fig.~\ref{fig:correctionFunction}, we multiply $\ell_m$ by the present outer correction factor $f(y/\delta)$:
  \begin{equation}
    \label{eq:ell_full_definition}
    \begin{split}
      \ell(y)&=\ell_m\,f\!\left(\frac{y}{\delta}\right)\,g(y^+)\\
            &=\kappa \alpha y\left(1-\frac{y}{\delta}\right)^{p}\left(1+\frac{y}{\delta}\right)^{q}\left(1-e^{-y^+/A}\right),
    \end{split}
  \end{equation}
  where $A$ is a configuration-dependent constant. With these definitions, the global eddy viscosity becomes a closed-form function of $y$:
  \begin{equation}
    \scriptsize
    \label{eq:nut_global_full}
    \begin{split}
      \frac{\nu_t}{\nu}
      =\frac{1}{2}\sqrt{1 + 4\left[
            \kappa \alpha y^+\left(1-\frac{y}{\delta}\right)^p\left(1+\frac{y}{\delta}\right)^q \left(1-e^{-y^+/A}\right)
        \right]^2\left(1-\frac{y}{\delta}\right)
      } -\frac{1}{2}.
    \end{split}
  \end{equation}
  Equivalently, absorbing the factor $(1-y/\delta)^{1/2}$ into the squared term yields the more compact form:
  \begin{equation}
    \footnotesize
    \label{eq:new_global_model}
    \begin{split}
      \frac{\nu_t}{\nu}
      =\frac{1}{2}\sqrt{1+4\left[\kappa \alpha y^+\left(1-\frac{y}{\delta}\right)^{p+1/2}\left(1+\frac{y}{\delta}\right)^q \left(1-e^{-y^+/A}\right)\right]^2}-\frac{1}{2}.
    \end{split}
  \end{equation}
  This expression retains the analytical structure of the Cess model, but replaces the Reichardt outer representation in Eq.~\eqref{eq:reichardt_nut_pipe} with the present correction function $f(y/\delta)$, thereby enabling a single full-depth formulation that remains accurate for closed-channel flow, open-channel flow, and pipe flow.
  We refer to Eq.~\eqref{eq:new_global_model} as the new global model, which may be viewed as a modified Cess model.
  The configuration-specific values of $A$ used in Eq.~\eqref{eq:new_global_model} are listed in Table~\ref{tab:outer_correction_params}.
  Figure~\ref{fig:eddymodelDNSCompares} compares the DNS eddy-viscosity distribution with the classical Cess model of Eq.~\eqref{eq:cess_model}, the DNS-based $k$--$\varepsilon$ estimate of Eq.~\eqref{eq:kepsilon_nut}, and the new model of Eq.~\eqref{eq:new_global_model}. Owing to the introduction of configuration-specific constants, the new model agrees better with the DNS data than the other global models, especially for open-channel flow.

\begin{figure*}[t]
\centering
\includegraphics[width=0.8\textwidth]{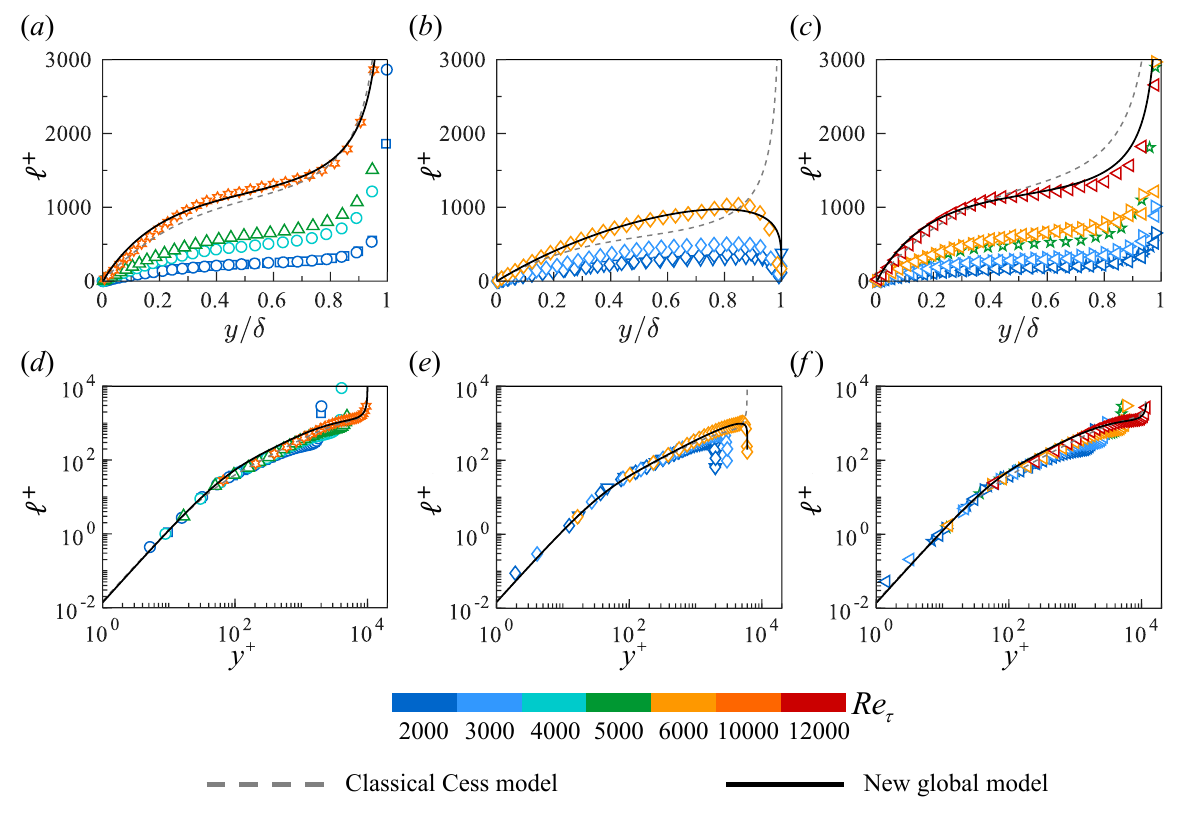} 
\caption{Wall-normal distributions of the characteristic length scale, $\ell^+=\ell/\delta_\nu$, for ((a), (d)) closed-channel flow, ((b), (e)) open-channel flow, and ((c), (f)) pipe flow at representative $Re_{\tau}$. Panels (a)-(c) show $\ell^+$ as a function of $y/\delta$ in linear coordinates, whereas panels (d)-(f) show the same quantity as a function of $y^+$ in semi-logarithmic coordinates. Symbols denote DNS-extracted characteristic length scales inferred from the mean-momentum relation. For clarity, the model curves are shown only for the highest-$Re_{\tau}$ case in each configuration, whereas the DNS results are shown for all cases.}
\label{fig:length_scale}
\end{figure*}

To provide a clearer physical interpretation of the proposed modifications, Fig.~\ref{fig:length_scale} compares the characteristic length scales in the three configurations.
For the DNS data, the characteristic length scale is extracted from the mean-momentum relation (Eq.~\eqref{eq:quadratic_for_S}), yielding
\begin{equation}
\ell_{\mathrm{DNS}}^{+}
=
\left[
\frac{1-y/\delta-S^{+}}{(S^{+})^{2}}
\right]^{1/2},
\end{equation}
in inner units, where $S^+=d\bar{u}^+/dy^+$. For the classical Cess model, based on Eq.~\eqref{eq:cess_model} together with the general eddy-viscosity form in Eq.~\eqref{eq:nut_nondim_general}, the implied characteristic length scale can be obtained analytically as
\begin{equation}
\ell_{\mathrm{Cess}}^{+}
=
\frac{\frac{\kappa Re_{\tau}}{3}
\left(1-z_c^2\right)\left(1+2z_c^2\right)
\left(1-e^{-z_c^{+}/A}\right)
}{
2\sqrt{1-y/\delta}
},
\end{equation}
where
\begin{equation}
z_c = 1-\frac{y}{\delta},
\qquad
z_c^{+} = Re_{\tau}\left(1-\frac{y}{\delta}\right),
\end{equation}
and $A$ is the damping constant.
For the present model, the corresponding length scale is given explicitly in Eq.~\eqref{eq:ell_full_definition}.
As shown in Fig.~\ref{fig:length_scale}, the DNS-inferred length scales exhibit clear configuration dependence in the outer region.
In closed-channel and pipe flows, $\ell^+$ increases away from the wall.
By contrast, in open-channel flow, $\ell^+$ decreases as the free surface is approached, reflecting a suppression of wake-like outer-layer buildup~\cite{pirozzoli2023loglaw}.
Constrained by its universal Reichardt-type outer representation, the classical Cess model fails to capture this physically expected downturn near the free surface and instead predicts an unphysical increase.
By contrast, the new global model, augmented by the outer correction function $f(y/\delta)$, captures these variations across the full depth or radius in all three flows.
Although the classical Cess model remains slightly more accurate in the canonical logarithmic region for closed-channel and pipe flows, the main advantage of the present model is its configuration-sensitive description of the outer-region length scale, which translates directly into improved predictions for open-channel flow.

\subsection{Effects of the eddy-viscosity model on mean-flow observables}
 We now assess the ability of the proposed eddy-viscosity model to represent the mean shear by examining the logarithmic indicator function~\cite{pirozzoli2023loglaw}.
  \begin{equation}
    \label{eq:loglaw_indicator_def}
    \Xi(y)=y^+\frac{d\overline{u}^+}{dy^+},
  \end{equation}
which provides a compact diagnostic of the mean-velocity gradient in viscous units. In an ideal logarithmic layer (Eq.~\eqref{eq:loglaw_nut_and_shear}), $\Xi=1/\kappa$ is constant, so departures from this plateau quantify deviations from logarithmic behavior. For eddy-viscosity-based closures, $\Xi$ can be evaluated directly from the total-stress relation (Eqs.~\ref{eq:total_shear_stress_linear} and~\ref{eq:compact_momentum_balance}), yielding
  \begin{equation}
    \label{eq:loglaw_indicator_from_nut}
    \Xi(y)=\frac{y^+\left(1-\frac{y}{\delta}\right)}{1+\frac{\nu_t}{\nu}}.
  \end{equation}
Hence, $\Xi$ can be computed from $\nu_t(y)$ for both the classical Cess model of Eq.~\eqref{eq:cess_model} and our present model of Eq.~\eqref{eq:new_global_model}. For Townsend's velocity-defect law~\cite{townsend1976}, $\Xi$ is obtained by differentiating the corresponding defect-law mean profile given by Eq.~\eqref{eq:outer_uplus_gradient} under the same nondimensionalization.

\begin{figure*}[t]
\centering
\includegraphics[scale=0.8]{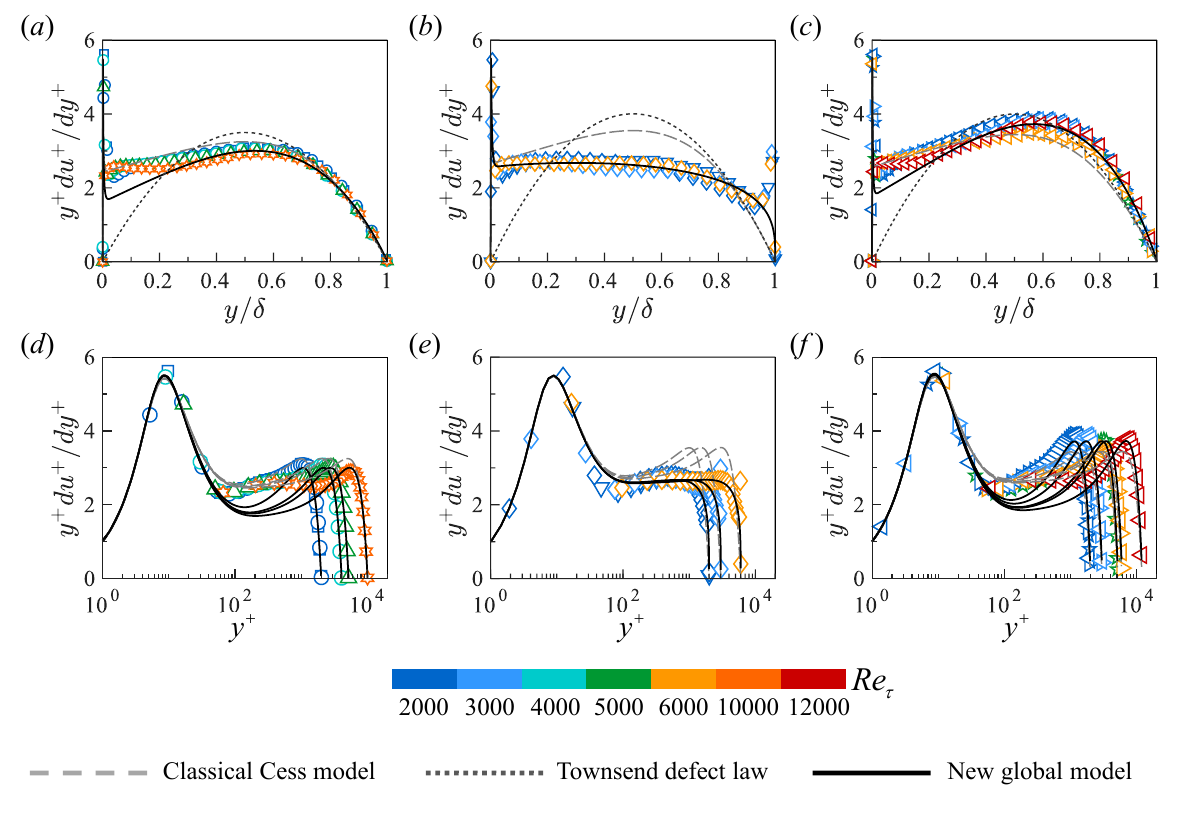}
\caption{Diagnostic assessment of the mean-velocity gradient via the log-law indicator function,
    $\Xi(y)\equiv y^{+}\,\mathrm{d}\overline{u}^{+}/\mathrm{d}y^{+}$, for ((a), (d)) closed-channel flow, ((b), (e)) open-channel flow, and ((c), (f)) pipe flow at representative values of $Re_{\tau}$. Panels (a)-(c) show $\Xi$ as a function of $y/\delta$ in linear coordinates, whereas panels (d)-(f) show the same quantity as a function of $y^{+}$ in semi-logarithmic coordinates. Symbols denote DNS results from Tables~\ref{tab:ccf_dns_datasets}-\ref{tab:pf_dns_datasets}. For clarity, in panels (a)-(c), the curves of the classical Cess model (Eq.~\eqref{eq:cess_model}) and the present model (Eq.~\eqref{eq:new_global_model}) are shown only for the highest Reynolds-number bin in each configuration, whereas DNS results are shown for all cases. Here, $\delta=h$ for closed-channel and open-channel flows and $\delta=R$ for pipe flow; in pipe flow, the wall distance is defined as $y=R-r$.}
\label{fig:diagnostic_function}       
\end{figure*}

Figure~\ref{fig:diagnostic_function}(a)-(c) shows $\Xi$ as a function of $y/\delta$ in linear coordinates.
Across closed-channel flow, open-channel flow, and pipe flow, Townsend's defect-law prediction~\cite{townsend1976} shows noticeable discrepancies in the outer region relative to the two eddy-viscosity models and does not reproduce the DNS trends consistently across the three configurations.
By contrast, both full-depth eddy-viscosity models (i.e., the classical Cess model and the present model) track the DNS-based indicator function more closely, indicating that enforcing the linear total-stress constraint through a full-depth $\nu_t(y)$ is important for predicting the outer-region shear.
The improvement provided by the present model is most pronounced in open-channel flow, whose outer-region behavior differs from that of closed-channel flow and pipe flow.
In closed-channel flow and pipe flow, the present model closely follows the classical Cess prediction. Because the outer-region shapes exhibit only weak Reynolds-number dependence within the selected bins, we plot the classical Cess and present-model curves only for the highest-$Re_\tau$ bin to avoid visual clutter, while showing the DNS results for all cases.

Figure~\ref{fig:diagnostic_function}(d)-(f) highlights the behavior of $\Xi$ in the near-wall and logarithmic regions by presenting it in semi-logarithmic coordinates.
Both eddy-viscosity-based models reproduce the expected near-wall increase and the emergence of an approximately constant $\Xi$ over an intermediate range of $y^+$, consistent with the presence of a logarithmic layer at sufficiently high Reynolds numbers.
However, as shown in Fig.~\ref{fig:diagnostic_function}(d) and (f) for closed-channel and pipe flows, the new global model is slightly less accurate than the optimized classical Cess model in the logarithmic region.
The origin of this discrepancy lies in the interaction between the extrapolated outer correction and the limited spatial reach of the near-wall damping function.
Because the proposed parametric outer correction is primarily optimized to capture the outer-region physics for $y/\delta\ge 0.20$, its inward extrapolation in symmetric flows exhibits a noticeable upward trend as $y/\delta\to 0$.
This behavior perturbs the pure $\kappa y$ scaling by increasing the implied length scale in the overlap region.
To maintain a robust full-depth velocity profile, the global fitting procedure therefore requires a compensatory increase in the near-wall damping constant $A$ (see Table~\ref{tab:outer_correction_params}).
Although this stronger damping improves the representation of the viscous sublayer and buffer layer, its effect is limited by the exponential form $\left(1-e^{-y^+/A}\right)$, which approaches unity in the logarithmic region.
Consequently, the damping function cannot fully suppress the mismatch that propagates inward from the extrapolated outer correction.
In open-channel flow, where the extrapolated outer correction remains monotonic, this issue is avoided, yielding a net improvement.
By contrast, in closed-channel and pipe flows, this interaction slightly perturbs the balance achieved by the classical Cess model.
This suggests that, although the present outer correction is essential for cross-configuration robustness, a more tailored inner damping function or a smoother inner--outer matching strategy may be needed in future work to optimize the logarithmic-region prediction fully.

\begin{figure*}[t]
\centering
\includegraphics[scale=0.8]{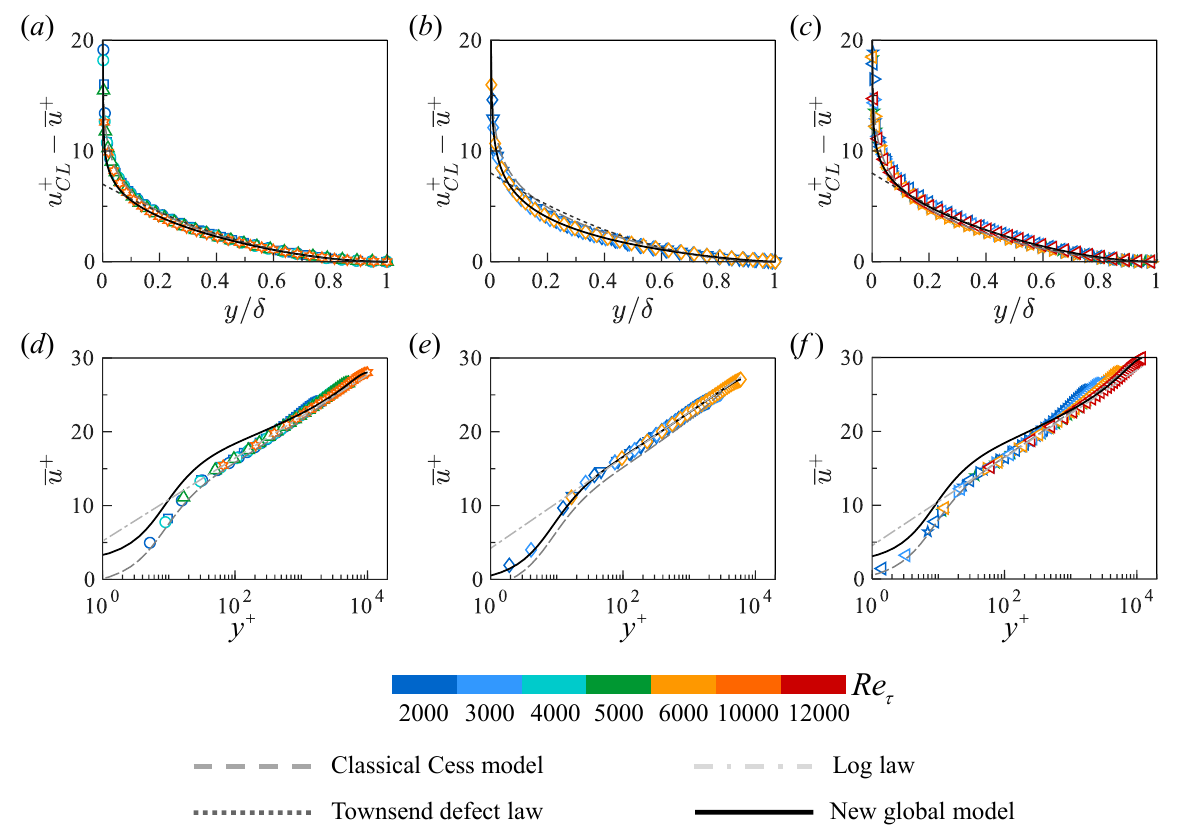}
    \caption{Mean-velocity profiles reconstructed from eddy-viscosity models and compared with DNS for ((a), (d)) closed-channel flow, ((b), (e)) open-channel flow, and ((c), (f)) pipe flow at representative values of $Re_{\tau}$.
      Panels (a)-(c) show the velocity-defect profiles $u_{CL}^{+}-\bar{u}^{+}$ as a function of $y/\delta$ in linear coordinates, whereas panels (d)-(f) show the corresponding mean-velocity profiles $\bar{u}^{+}$ as a function of $y^{+}$ in semi-logarithmic coordinates.
      Symbols denote DNS results from Tables~\ref{tab:ccf_dns_datasets}-\ref{tab:pf_dns_datasets}. The model-based mean-velocity profiles are obtained by numerically integrating the predicted mean shear and are aligned by matching the centerline or free-surface velocity.
      For clarity, in panels (a)-(c), the curves of the classical Cess model (Eq.~\eqref{eq:cess_model}) and the present model (Eq.~\eqref{eq:new_global_model}) are shown only for the highest Reynolds-number bin in each configuration, whereas DNS results are shown for all cases.}
\label{fig:velocityprofile}       
\end{figure*}

We next assess the performance of our eddy-viscosity model in predicting the mean velocity. For a prescribed eddy-viscosity distribution $\nu_t(y)$, the total-stress relation yields the mean shear,
  \begin{equation}
    \frac{\mathrm{d}\bar{u}}{\mathrm{d}y}
    =
    \frac{u_\tau^{2}\left(1-\dfrac{y}{\delta}\right)}{\nu+\nu_t(y)}.
    \label{eq:mean_shear}
  \end{equation}
Because the resulting expression is algebraically cumbersome for both the classical Cess model and the present model, we do not pursue an analytical closed-form expression for $\bar{u}(y)$. Instead, we reconstruct the mean velocity numerically as
  \begin{equation}
    \bar{u}(y)
    =
    \int_{0}^{y}\frac{\mathrm{d}\bar{u}}{\mathrm{d}y'}\,\mathrm{d}y' + C,
    \label{eq:mean_velocity_reconstruction}
  \end{equation}
where $y$ is the wall-normal distance from the nearest solid wall, $y'$ is a dummy integration variable, and $C$ is an integration constant.
For the DNS data, $C=0$ naturally. For the model-based reconstructions, however, small discrepancies in the predicted mean shear inevitably accumulate during integration.
Because our primary interest lies in the outer-region behavior, we choose $C$ such that the reconstructed profile matches a reference outer velocity $u_{CL}^{+}$, namely the centerline velocity in closed-channel and pipe flows and the free-surface velocity in open-channel flow.
This choice is consistent with the velocity-defect representation in Fig.~\ref{fig:velocityprofile}(a)-(c), in which the defect is defined as $u_{CL}^{+}-\bar{u}^{+}$~\cite{pirozzoli2023loglaw}.

Figure~\ref{fig:velocityprofile}(a)-(c) shows that, in the outer region, the present model reproduces the DNS velocity defect more accurately than Townsend's defect law (Eq.~\eqref{eq:quadratic_defect}). The improvement is particularly clear in open-channel flow, consistent with the configuration-dependent outer correction introduced in Fig.~\ref{fig:correctionFunction}.
In closed-channel and pipe flows, the present model yields defect profiles close to those obtained with the classical Cess model, indicating that the main benefit of the present modification is to regularize the outer-region behavior across different boundary conditions rather than to alter the bulk defect scaling in these two configurations.
Although Townsend's defect law~\cite{townsend1976} relies on the simplifying assumption of an approximately uniform eddy viscosity in the outer region, it remains practically applicable to closed-channel and pipe flows.
This helps explain its continued popularity as a useful outer-region reference despite its limited interpretability as a full-depth closure. A similar observation applies to the classical Cess model in open-channel flow.

Figure~\ref{fig:velocityprofile}(d)-(f) presents the mean-velocity profiles in semi-logarithmic coordinates. In open-channel flow, the present model remains close to the DNS data over a broad wall-normal range, and the apparent logarithmic region is comparatively long.
In closed-channel and pipe flows, the present model exhibits a modest mismatch in the logarithmic region relative to DNS, consistent with its slightly less accurate prediction of the indicator function $\Xi=y^{+}\,\mathrm{d}\bar{u}^{+}/\mathrm{d}y^{+}$ in Fig.~\ref{fig:diagnostic_function}(d) and (f).
By contrast, the classical Cess model (Eq.~\eqref{eq:cess_model}) performs slightly better in the logarithmic region for pipe flow. As a result, the mean-velocity profile predicted by the present model is consistently higher than that predicted by the classical Cess model in the near-wall region.
This behavior suggests that the remaining discrepancy is primarily associated with the near-wall damping/blending inherited from the Cess-type framework rather than with the outer correction itself. It therefore points to a possible route for further improvement, namely refining the near-wall treatment while preserving the outer-region robustness achieved here.
Among the three configurations, open-channel flow exhibits the longest apparent logarithmic region, which may be related to a weaker wake component, as discussed by Pirozzoli~\cite{pirozzoli2023loglaw}.
This observation is consistent with the improved full-depth agreement of the present model in open-channel flow and supports the view that the outer boundary condition, namely the free-slip surface, plays an important role in shaping the outer eddy viscosity and, consequently, the reconstructed mean velocity.

We next discuss the Reynolds-number sensitivity and predictive robustness of these parameters. An \textit{a posteriori} bin-wise fitting analysis shows that, whereas the parameters for open-channel flow are relatively stable, those for closed-channel and pipe flows (particularly $\alpha$ and $A$) retain a mild dependence on the friction Reynolds number.
These parameters are therefore not strictly Reynolds-number independent.
To assess the predictive robustness of using a single global parameter set (Table~\ref{tab:outer_correction_params}) rather than $Re_\tau$-specific parameters, we quantified the weighted root-mean-square (RMS) errors of the reconstructed mean-velocity profiles.
The analysis (not shown here for brevity) indicates that, although using a single parameter set slightly increases the local error in the eddy-viscosity distribution, the resulting RMS errors in the integrated mean velocity $\bar{u}^+$ remain essentially identical to those obtained with $Re_\tau$-specific fits over the full Reynolds-number range considered here.
This result demonstrates that a single, configuration-specific parameter set provides a robust predictive baseline for high-Reynolds-number mean-flow reconstruction without requiring dynamic $Re_\tau$-dependent tuning.
It is also useful to clarify the theoretical behavior of the proposed model near the outer boundary, $y=\delta$.
In the present model (Eq.~\eqref{eq:new_global_model}), the asymptotic scaling near $y=\delta$ is $\nu_t\propto(1-y/\delta)^{p+0.5}$.
For open-channel flow, the fitted value $p=0.20$ yields a positive exponent, ensuring that the modeled eddy viscosity vanishes smoothly at the free surface, consistent with the free-slip boundary condition.
For closed-channel and pipe flows, a strictly finite centerline value would require $p=-0.5$.
Our unconstrained fits yield $p=-0.46$ and $p=-0.43$, respectively. Because these values are slightly larger than $-0.5$, the modeled $\nu_t$ formally vanishes at $y=\delta$.
However, because the exponent $p+0.5$ remains very small (approximately $0.04$ and $0.07$), the model retains a near-finite plateau that matches the DNS trend up to $y/\delta\approx0.99$, thereby providing a practically robust approximation without artificially overconstraining the global fit.

\begin{figure*}[t]
\centering
\includegraphics[scale=0.8]{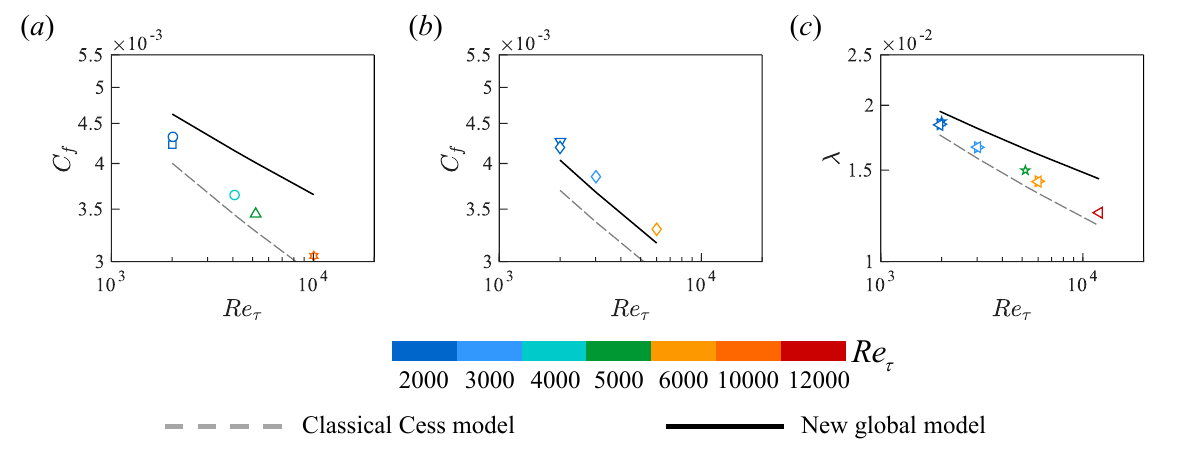}
    \caption{Skin-friction measures as functions of the friction Reynolds number for pressure-driven wall turbulence in (a) closed-channel flow, (b) open-channel flow, and (c) pipe flow. Panels (a) and (b) show the skin-friction coefficient $C_f=\frac{2\tau_w}{\rho\,u_b^2}$ for closed-channel and open-channel flows, whereas panel (c) shows the friction factor $\lambda=\frac{8\tau_w}{\rho\,u_b^2}$ for pipe flow. Symbols denote DNS results from Tables~\ref{tab:ccf_dns_datasets}-\ref{tab:pf_dns_datasets}.}
\label{fig:Cf}       
\end{figure*}

We finally assess the predictive performance of the eddy-viscosity model for skin friction, an integral quantity of practical engineering interest. For closed-channel and open-channel flows, we define the skin-friction coefficient~\cite{xia2021skin} based on the bulk velocity $u_b$ as
  \begin{equation}
    C_f=\frac{2\tau_w}{\rho\,u_b^2}.
    \label{eq:skin_friction_Cf}
  \end{equation}
For pipe flow, we report the friction factor~\cite{yao2023pipe},
  \begin{equation}
    \lambda=\frac{8\tau_w}{\rho\,u_b^2}.
    \label{eq:pipe_friction_factor}
  \end{equation}
For the DNS data, $u_b$ and $\tau_w$ are taken directly from the reported database statistics. For the classical Cess model (Eq.~\eqref{eq:cess_model}) and the present model (Eq.~\eqref{eq:new_global_model}), the bulk velocity is defined as the cross-sectional average. Figure~\ref{fig:Cf} compares the corresponding friction quantities obtained from DNS with those predicted by the models.
The present model provides the clearest improvement for open-channel flow, where the outer eddy-viscosity trend differs from those of closed-channel and pipe flows.
In closed-channel and pipe flows, the present model yields friction levels close to those predicted by the classical Cess model.
We also observe a slight overprediction of the friction level in closed-channel and pipe flows relative to both DNS and the classical Cess model. This trend is consistent with the slight deterioration of the logarithmic-region shear (see Fig.~\ref{fig:diagnostic_function}), which affects the bulk-velocity integral and, consequently, the inferred friction level.

\section{Conclusion}\label{sec:section5}
In this study, we revisited eddy-viscosity distributions in high-Reynolds-number pressure-driven wall turbulence from a comparative perspective across configurations, focusing on closed-channel flow, open-channel flow with a free-slip surface, and pipe flow.
Using DNS one-point statistics, namely the mean velocity and Reynolds shear stress, together with the exact mean-momentum balance, we inferred the eddy-viscosity distribution $\nu_t$ over the full depth or radius and identified configuration-dependent trends in the outer region that cannot be captured by a single conventional full-depth expression.

Interpreting $\nu_t$ as the product of a velocity scale and a length scale, we extended this viewpoint into the outer region by defining a local velocity scale from the linear total-stress distribution, $u_c\approx{}u_\tau\sqrt{1-y/\delta}$, while retaining the mixing-length form $\ell_m=\kappa\,y$.
The remaining configuration dependence was isolated in an outer correction factor $f$, expressed as a function of $y/\delta$, which was extracted from DNS and found to be only weakly sensitive to Reynolds number within the selected bins, while differing among closed-channel, open-channel, and pipe flows.
We proposed a compact parametric representation of the outer correction $\kappa\,f$ as a function of $y/\delta$ that accommodates both monotonic outer trends, characteristic of open-channel flow, and non-monotonic ones, characteristic of closed-channel and pipe flows, through the exponents $p$ and $q$, thereby providing a minimal yet flexible outer ingredient for global modeling.
We then embedded the fitted outer correction into a Cess-type analytical framework by combining it with a van Driest near-wall damping function, yielding an explicit full-depth eddy-viscosity model, here referred to as the new global model.
The resulting model preserves the algebraic convenience of classical Cess-type constructions while improving robustness across configurations through the outer correction.

The impact of different eddy-viscosity representations on mean-flow observables was assessed through (1) the logarithmic indicator function $\Xi=y^{+}\,\mathrm{d}\bar{u}^{+}/\mathrm{d}y^{+}$, (2) reconstructed mean-velocity and velocity-defect profiles, and (3) skin friction.
The present model yields the most pronounced improvement in open-channel flow, where the outer eddy-viscosity trend differs most strongly from those in closed-channel and pipe flows, leading to better agreement in the outer-region shear, velocity-defect profile, and friction prediction.
In closed-channel and pipe flows, the present model produces results close to those of the classical Cess model, indicating that the main benefit of the proposed modification is to regularize the outer-region behavior across different boundary conditions rather than to alter the overall scaling in configurations where the classical Cess model already performs well.
The remaining discrepancies are primarily associated with the near-wall and logarithmic regions, reflecting limitations of the current blending between the van Driest damping and the outer correction within a single closed-form expression.
Future improvements may therefore focus on refining the near-wall damping or blending strategy while preserving the present outer correction, which provides robustness across closed-channel, open-channel, and pipe flows.

\vspace{0.5\baselineskip}
\noindent {\footnotesize \bf Conflict of interest} {\footnotesize \it \quad On behalf of all authors, the corresponding author states that there is no conflict of interest.}\par

\vspace{0.5\baselineskip}
\noindent {\footnotesize \bf Author contributions} {\footnotesize \it \quad \textbf{Ben-Rui Xu}: Conceptualization, Methodology, Formal analysis, Investigation, Writing -- original draft.
\textbf{Ao Xu}: Conceptualization, Methodology, Supervision, Funding acquisition, Resources, Writing -- review \& editing. }\par
\Acknowledgements{{\bf Acknowledgements} \quad This work was supported by the National Natural Science Foundation of China (NSFC) (Grant Nos. 12388101 and 12272311), the Young Elite Scientists Sponsorship Program by CAST (Grant No. 2023QNRC001), and the Fundamental Research Funds for the Central Universities (Grant No. D5000260269). The authors acknowledge the Computing Center in Xi’an for providing HPC resources that have contributed to the research results reported within this paper.}

\end{multicols}

\entitle{再探压力驱动高雷诺数壁湍流中的涡黏系数}{再探压力驱动高雷诺数壁湍流中的涡黏系数}
\enauthor{许本瑞, 徐翱}{}
\enabstract{本文研究了三类典型压力驱动壁湍流中的涡黏系数分布, 包括封闭槽道流、具有自由滑移表面的开放槽道流以及圆管流. 基于覆盖摩擦雷诺数$Re_\tau = 2000$-12000的直接数值模拟数据库, 我们通过Boussinesq关系, 利用单点统计量反推出涡黏系数. DNS反推结果表明, 涡黏系数在外区呈现明显的构型依赖特征, 说明单一的全高度表达式难以在三类构型中均保持一致的预测精度. 基于涡黏系数可表示为速度尺度与长度尺度乘积的物理图像, 本文将对数律标度推广至外区. 具体而言, 本文采用基于应力的速度尺度, 并进一步引入外区修正函数, 以表征外区尺度效应. 随后, 将该修正函数的参数化形式嵌入包含van Driest近壁阻尼的Cess型框架中, 建立了一个适用于全高度范围的涡黏系数分布模型. 我们进一步通过涡黏系数剖面、对数律诊断函数以及壁面摩擦系数对该模型进行评估. 结果表明, 所提出的模型在开放槽道流中具有明显改进, 同时在封闭槽道流和圆管流中仍保持与经典Cess模型相当的预测精度. 上述结果强调了外边界条件对外区涡黏系数分布的调控作用, 并进一步揭示了其对平均流预测的影响.}

\noindent\makeentitle

\end{document}